\documentclass{aastex}
\usepackage{spr-astr-addons}
\usepackage{url}
\usepackage{graphicx,times}
\usepackage{bm}
\usepackage{epsfig}
\usepackage{amsmath,amsfonts}
\usepackage{url}

\RequirePackage{color}

\begin{document}

\title{Local stability of self-gravitating disks in $f(R)$ gravity}
\slugcomment{Not to appear in Nonlearned J., 45.}
%% Running heads
\shorttitle{Short article title}
\shortauthors{Autors et al.}

\author{Mahmood Roshan\altaffilmark{1}} 
\affil{mroshan@um.ac.ir}
\and
\author{Shahram Abbassi\altaffilmark{1,2}}
\affil{abbassi@um.ac.ir}

\altaffiltext{1}{Department of Physics, Ferdowsi University of Mashhad, P.O. Box 1436, Mashhad, Iran}
\altaffiltext{2}{School of Astronomy, Institute for Research in Fundamental Sciences (IPM), P.O. Box 19395-5531, Tehran, Iran}
\begin{abstract}
In the framework of metric $f(R)$ gravity, we find the dispersion relation for the propagation of tightly wound spiral density waves in the surface of rotating, self-gravitating disks. Also, new Toomre-like stability criteria for differentially rotating disks has been derived for both fluid and stellar disks. 
\end{abstract}
 \section{Introduction}
The stars, gas and dust clouds in galactic disks congregate in spiral patterns and make  bountiful structures in the universe. Despite spiral arms beauties and many decades of concerted scientific investigation, much about them remain mysterious. Although the precise theory explaining the origin and evolution of the spiral structures in spiral galaxies is not fully understood, it is widely agreed in the relevant literature that these patterns are gravitationally driven density waves in the stellar disks \citep{binney,sellwood}. Therefore, the \textit{density wave theory} is an important tool for studying the dynamics and the evolution of spiral galaxies. In the frame work of this theory and using some approximations and assumptions, Alar Toomre showed that the differentially rotating disks in Newtonian dynamics is stable to all local axisymmetric disturbances if the dimensionless quantity $Q>1$ \citep{toomre}, where $Q$ is the Toomre's $Q$ parameter (see equations (\ref{Q1}) and (\ref{Q2}) for the definition of this parameter for fluid and stellar disks respectively). When $Q<1$, on the other hand, thermal pressure and rotation are unable to stop the collapse of over-dense regions.

In fact, he assumed that the density waves are tightly wound and the stellar orbits of the unperturbed disk are nearly circular. Despite these restrictive assumptions used in deriving this result, it has proved to be remarkably true and widely applicable. However, numerical studies of disk galaxies reveals that this criterion is not enough for complete stability of the stellar self-gravitating disks \citep{nbody1,nbody2, peebles}. In other words, these N-body simulations showed that Toomre's criterion does not provide the global stability of the stellar disks and only guarantees the local stability of them. More specifically, it turned out that simple models of rotationally dominated stellar disks are globally unstable to a pressure dominated bar-like structure. 

This kind of gravitational instability can be directly linked to the dark matter problem in spiral galaxies. This link has been known since 1973 when Ostriker and Peebles pointed out that a dark matter halo surrounded the galaxy may be an important stabilizing agent of galactic disks \citep{peebles} and prevent the rapid bar formation. However, it should be noted that existence of a spherical halo around a rotationally dominated (or cold) disk is not the only way to stabilize the disk. There are other possibilities which can provide the global stability, see \citep{sellwood} for a review of the subject. For example, existence of a massive central bulge would stabilize the disk \citep{sellwood2,sellwood3,sellwood4}.  

The main aim of this paper is to find the generalized Toomre's local stability criterion in the context of metric $f(R)$ gravity. As an example of such a study in other modified gravity theories see \citep{mil1} where the Toomre's criterion has been derived in the context of modified Newtonian dynamics (MOND). Also, \cite{us3} have derived the Toomre's stability criterion in the context of MOG. It is worth to mention that MOG is a Scalar-Tensor-Vector theory of gravity presented to address the dark matter problem \citep{moffat}. Although a considerable amount of work has gone into stability criteria, no study has been performed for investigating these criteria in the context of $f(R)$ gravity. However, it should be noted that the dynamical stability of spherical systems, and also the stability of spherically symmetric solutions  in $f(R)$ gravity have been widely investigated (see \cite{r1} and \cite{r2} and references therein for more details.) Certainly, the results of this paper could be useful in the numerical simulations of the disk galaxies in $f(R)$ gravity.

%we think that it is important to derive such a widely useful formulation.

It is worth mentioning that metric $f(R)$ gravity is one of the simplest modifications to Einstein's General Relativity (GR). Strictly speaking, the generic action of this theory can be simply obtained just by replacing the Ricci scalar $R$ with an arbitrary general function of $R$, namely $f(R)$, in the Einstein-Hilbert action. Several aspect of this theory have been investigated in the literature. Interests to this theory increased when Carroll \textit{et al} proposed that $f(R)$ gravity can solve the cosmic speedup enigma \cite{carroll}. Although this theory is known as a dark energy model, it has been applied to address the dark matter problem in the galactic scales. For example \cite{r3} applied a power-law $f(R)$ gravity to explain the rotation curves of low surface brightness galaxies. We refer the reader to review papers \citep{faraoni,clifton} for more detail about this theory and its status among the other extended theories of dark energy.  

The structure of this paper is as follows. The weak field approximation of metric $f(R)$ gravity is reviewed briefly in section \ref{Newton}. Also, the coupled Boltzmann and modified Poisson equations are derived.  In section \ref{fluid}, by linearizing the field equations, we derive the dispersion relation of tightly wound spiral density waves in both fluid and stellar disks. Furthermore, using the new dispersion relations, we derive the local stability criterion. In fact, we find the generalized version of the Toomre's stability criterion in the context of metric $f(R)$ gravity. This new Toomre's criterion is the main result of this paper. Finally, in section \ref{DC}, results are discussed.

\section{Weak field limit of $f(R)$ gravity}\label{Newton}
Here, we briefly review the weak filed limit of $f(R)$ gravity theory. The weak field limit of this theory has been widely investigated, for a comprehensive review of the subject see \citep{faraoni,fara1,fara2,fara3}. The general action for this theory is given by
\begin{eqnarray}
S=\frac{1}{16\pi G}\int\sqrt{-g} f(R)d^4x+S_M
\label{action}
\end{eqnarray}
where $R, G, g, S_M$ are the Ricci scalar, Newton's gravitational constant, the determinant of the metric tensor and  the matter action, respectively. Note that throughout this paper we work in the system of units where the speed of light is $c=1$. Furthermore, we will assume that $f(R)$ is analytic
about $R = 0$ so that it can be expanded as a power series 
\begin{eqnarray}
f(R)=\sum c_n R^n
\label{model}
\end{eqnarray}
There are some $f(R)$ models of cosmological interest in the literature that can be expressed in such a form, for example, Starobinsky's model \cite{staro} 
\begin{eqnarray}
f(R)=R+\beta R_0 \left[\left(1+\frac{R^2}{R_0^2}\right)^{-m}-1\right]
\end{eqnarray}
can be expressed as
\begin{eqnarray}
f(R)=R-\frac{m\beta}{R_0}R^2+\frac{\beta m (m+1)}{2R_0^3}R^4+...
\end{eqnarray}
Variation of action (\ref{action}) with respect to metric yield to the metric field equation:
\begin{equation}
f'R_{\mu\nu}-\frac{f}{2}g_{\mu\nu}-\nabla_{\mu}\nabla_{\nu}f'g_{\mu\nu} \Box f'=8\pi G T_{\mu\nu}
\label{fe}
\end{equation}
where prime denotes derivative with respect to $R$, i.e. $f'(R)=\frac{df}{dR}$, and $T_{\mu\nu}$ is the energy-momentum tensor. Throughout this paper, we restrict ourselves to the adiabatic approximation in which the evolution of the universe is very slow in comparison with local dynamics. This assumption allows us to use the Minkowski space-time instead of more natural choices such as the Friedmann-Robertson-Walker (FRW) space-time as the background metric (see \citep{clifton} and references cited therein). It should be noted that in the current paper we consider the local gravitational stability of the self-gravitating disks. This means that only the local dynamics of these systems matters for us.

Therefore, in order to write the field equations in the first perturbation order, we decompose metric into the Minkowski metric $\eta_{\mu\nu}=\text{diag}(+1,-1,-1,-1)$ plus a small perturbation $h_{\mu\nu}$ ($|h_{\mu\nu}|\ll 1$) as follows
\begin{eqnarray}
ds^2=(1+2 \Phi)dt^2-(1-2\Psi)(dx^2+dy^2+dz^2)
\label{metric}
\end{eqnarray}
where $h_{00}=2\Phi(\mathbf{x},t)$ and $h_{ij}=2\Psi(\mathbf{x},t)\delta_{ij}$. We use the \textit{transverse gauge}, i.e. $\partial_i s^{ij}=0$.
where the traceless tensor $s_{ij}$ is known as the \textit{strain} and is given by
\begin{eqnarray}
s_{ij}=\frac{1}{2}\left(h_{ij}-\frac{1}{3}\delta^{kl}h_{kl}\delta_{ij}\right)
\end{eqnarray}
Finally, substituting (\ref{metric}) into field equation (\ref{fe}) and keeping only terms linear in perturbations $\Phi$ and $\Psi$, one can find the following equation for the $(0,0)$ component of the field equation
\begin{eqnarray}
f''_0\nabla^4(2\Psi-\Phi)+f'_0 \nabla^2\Psi=8\pi G \rho
\label{main1}
\end{eqnarray}
where $f^{(n)}_0=\frac{d^n f}{dR^n}|_{R=0}$ and $\nabla^4 \psi=\nabla^2(\nabla^2\psi)$. Similarly, preserving only the first-order perturbations, the trace of the field equation (\ref{fe}) takes the following form
\begin{eqnarray}
3f''_0\nabla^4(2\Psi-\Phi)+f'_0 \nabla^2(2\Psi-\Phi)=8\pi G \rho
\label{main2}
\end{eqnarray} 

It must be mentioned that we have assumed the perfect fluid's energy-momentum tensor for $T_{\mu\nu}$. Thus $\rho$ is the rest frame matter density. Equations (\ref{main1}) and (\ref{main2}) all together with the continuity equation and the Euler, make a complete set of equations governing the dynamics of a fluid system in the weak filed limit of $f(R)$ gravity. In fact, equations (\ref{main1}) and (\ref{main2}) are the analog of the Poisson equation in Newtonian gravity. It is obvious that we have to find the corresponding equations for  continuity equation and the Euler equation in the frame work of $f(R)$ gravity. To do so, it is needed to take into account that metric $f(R)$ theory is a metric theory of gravity. It is known that in metric theories of gravity the conservation of the energy momentum-tensor holds, i.e. $\nabla_{\mu}T^{\mu\nu}=0$. Consequently, non-spinning test particles move on the geodesics of the metric tensor $g_{\mu\nu}$. Therefore, the equation of motion of a test particle in this theory is given by
\begin{eqnarray}
\frac{d^2x^i}{d\tau^2}+\Gamma^{i}_{\alpha\beta}\frac{dx^{\alpha}}{d\tau}\frac{dx^{\beta}}{d\tau}=0
\label{geo}
\end{eqnarray}
where $\tau$ is the proper time along the world line of the test particle. Using the metric component (\ref{metric}), it is straightforward to write Eq. (\ref{geo}) to first order in perturbations
\begin{eqnarray}
\frac{d^2\mathbf{r}}{dt^2}=-\nabla \Phi
\label{new0}
\end{eqnarray} 
thus, only the metric potential $\Phi$ directly appears in the equation of motion of a test particle. In other words, only $\Phi$ appears in the gravitational potential. This is the case also for the continuity and  Euler equation. Writing
the divergence of the energy-momentum tensor, $\nabla_{\mu}T^{\mu\nu}=0$, in the Newtonian limit, one can easily verify that the continuity and Euler equations are
\begin{eqnarray}
\frac{\partial \rho}{\partial t}+\nabla\cdot (\rho \mathbf{v})=0
\label{conti}
\end{eqnarray}
\begin{eqnarray}
\frac{\partial \mathbf{v}}{\partial t}+(\mathbf{v}\cdot \nabla)\mathbf{v}=-\frac{1}{\rho}\nabla p-\nabla\Phi
\label{eul}
\end{eqnarray}
Equations (\ref{main1}),(\ref{main2}), (\ref{conti}) and (\ref{eul}) together with the equation of state relating $p$ and $\rho$, make a set of equations governing the dynamics of a self-gravitating fluid system.

\subsection{The modified Poisson equation in $f(R)$ gravity}
In this section we derive the Poisson equation in the weak field limit of metric  $f(R)$ gravity. For this purpose, it is convenient to combine equations (\ref{main1}) and (\ref{main2}) as a single equation. The result is the analouge the Poisson equation in Newtonian gravity. Using equations (\ref{main1}) and (\ref{main2}), one can show that
\begin{eqnarray}
\nabla^2\Psi=-\nabla^2 \Phi+\frac{8\pi G}{f'_0}\rho
\label{ps}
\end{eqnarray}
Now, substituting equation (\ref{ps}) into (\ref{main2}), we find
\begin{eqnarray}
(\nabla^2-m_0^2)\nabla^2\Phi=-4\pi G \alpha \left(m_0^2 \rho-\frac{4}{3 }\nabla^2\rho\right)
\label{po}
\end{eqnarray}
where $\alpha=1/f'_0$ and $m_0$ is defined as
\begin{eqnarray}
m_0^2=-\frac{f'_0}{3 f''_0}
\end{eqnarray}
in fact, equation (\ref{po}) is the modified version of the Poisson equation. Therefore, loosely speaking, we are dealing with a theory with two free parameters $\alpha$ and $\beta$. In the limit $m_0^2\rightarrow \infty$ and $\alpha\rightarrow 1$, the standard Poisson equation is recovered, i.e. equation (\ref{po}) takes the form $\nabla^2\Phi=4\pi G \rho$. Solution of this fourth order equation (\ref{po}) can be written as follows
\begin{eqnarray}
\begin{split}
\Phi(\mathbf{r})=-4\pi G\alpha &\int\int G_1(\mathbf{r}'',\mathbf{r}')G_2(\mathbf{r},\mathbf{r}'')\\&\times\left(m_0^2 \rho(\mathbf{r}')-\frac{4}{3 }\nabla^2\rho(\mathbf{r}')\right)d^3\mathbf{r}'d^3\mathbf{r}''
\label{po1}
\end{split}
\end{eqnarray}
where $G_1$ and $G_2$ are the Green functions:
\begin{eqnarray}
\begin{split}
& G_1(\mathbf{r}'',\mathbf{r}')=-\frac{1}{4\pi}\frac{e^{-\sqrt{m_0^2}|\mathbf{r}'-\mathbf{r}''|}}{|\mathbf{r}'-\mathbf{r}''|}\\& G_1(\mathbf{r},\mathbf{r}'')=-\frac{1}{4\pi}\frac{1}{|\mathbf{r}-\mathbf{r}''|}
 \end{split}
\end{eqnarray}
it is natural to expect that the free parameters $\alpha$ and $m_0$ should be fixed by relevant observations.  Also, in order to prevent the oscillatory solutions, $m_0^2$ should be positive, i.e. $m_0^2>0$.

 It is worth noting that the gravitational potential of a point source takes the following form
\begin{eqnarray}
\Phi(r)=-\frac{G M \alpha}{r}\left[1+\frac{1}{3}e^{-m_0 r}\right]
\label{psou}
\end{eqnarray}
This result can be straightforwardly derived from equation (\ref{po1}) just by substituting the matter density as $\rho(\mathbf{r})=M \delta(\mathbf{r})$, where $M$ is the mass of the point particle.
\subsection{The collisionless Boltzmann equation in $f(R)$ gravity}\label{bolbol}
In this section, we derive the collisionless Boltzamnn equation in the weak field limit of metric $f(R)$ gravity. This equation is the main equation governing the dynamics of a stellar system. Assume that $f(x^{\mu}, u^{\mu})$
is the particle's phase space distribution function. Where $u^{\mu}=\frac{dx^{\mu}}{d\tau}$ is the four velocity of the particle. In Newtonian dynamics the Boltzmann equation can be expressed as $\hat{L}[f]=0$, where $\hat{L}$ is the Liouville operator.
It is worth remembering that in metric $f(R)$ gravity non-spinning test particles move on the geodesics. Therefore with the aid of the geodesic equation, we obtain the covariant, relativistic generalization of the Boltzmann equation
\begin{eqnarray}
\hat{L}[f]=u^{\mu}\frac{\partial f}{\partial x^{\mu}} -\Gamma^{\mu}_{\alpha\beta}u^{\alpha}u^{\beta}\frac{\partial f}{\partial u^{\mu}}=0
\end{eqnarray}
It is straightforward to linearize this equation using the perturbation introduced in Eq. (\ref{metric}). Keeping in mind that in the first order Newtonian limit
$\Gamma^i_{00}\simeq \frac{\partial\Phi}{\partial x^i}$ and $u^i\simeq v^i=\frac{dx^i}{dt}$, we have
\begin{eqnarray}
\frac{\partial f}{\partial t}+ \mathbf{v}\cdot\nabla f-\nabla \Phi\cdot\frac{\partial f}{\partial \mathbf{v}}=0
\label{bolt}
\end{eqnarray}
as expected, only metric potential $\Phi$ appears in the Boltzmann equation. Mathematically, this equation is the same as the Boltzmann equation in Newtonian dynamics. However, we know that the field equation governing $\Phi$ is different from the standard Poisson equation. 

\section{Local stability of differentially rotating disks in $f(R)$ gravity}\label{fluid}

As we mentioned in the introduction, our main purpose in this paper is to find a Toomre-like stability criterion in the context of metric $f(R)$ gravity. We remind the reader that Toomre's criterion for a fluid (gaseous) disk is expressed as
\begin{eqnarray}
Q_g=\frac{v_s \kappa}{\pi G\Sigma}>1
\label{Q1}
\end{eqnarray} 
where $v_s$ is the sound speed in the fluid, $\kappa$ is the epicycle frequency and $\Sigma$ is the surface matter density of the disk. Similarly, in the case of a stellar disk this criterion can be written as 
\begin{eqnarray}
Q_s=\frac{\sigma \kappa}{3.36 G\Sigma}>1
\label{Q2}
\end{eqnarray} 
in this case $\sigma$ is the radial velocity dispersion.
In Newtonian gravity, Toomre's criterion has been derived for differentially rotating disk where the \textit{tight-winding} or the \textit{WKB} approximation is satisfied. In fact, the tight-winding (or WKB) approximation provides some simplifications in the story and, without much loss of generality of the problem,  makes analytic description possible and much simpler. In this approximation, the density wave in the disk can locally be regarded as a plane wave. This simplification enables us to find the dispersion relation for this kind of density waves and to describe their propagation in the disk. Therefore, WKB approximation is an indispensable tool for understanding the behavior of density waves in rotating disks in the context of Newtonian gravity. As we shall see in the next sections, in order to do complete stability analysis in $f(R)$ gravity and find the generalized version of the Toomre's criterion, we also need to benefit this approximation.

\subsection{Self-gravitating fluid disk}\label{self fluid}

In this section, we analyze the behavior of a tightly wound density wave in the framework of metric $f(R)$ gravity. First, we use the modified Poisson equation (\ref{po}) in order to calculate the gravitational potential of a tightly wound surface density. Furthermore, we linearize the relevant equations in order to find the dispersion relation for such a density wave.

The continuity equation (\ref{conti}), the Euler equation (\ref{eul}) and the modified Poisson equation (\ref{po}) are the required equation for studying the gravitational stability of the system. It is important to remember that we also need the equation of state of the fluid to make a complete description. We assume that the background disk is axisymmetric and is placed in $z = 0$ plane. Using the cylindrical coordinates $(R,\phi, z)$, the continuity equation reads (note that hereafter $R$ is a coordinate and should not be confused with the Ricci scalar)
\begin{eqnarray}
\frac{\partial\Sigma}{\partial t}+\frac{1}{R}\frac{\partial}{\partial R}\left(\Sigma R v_{R}\right)+\frac{1}{R}\frac{\partial}{\partial\phi}\left(\Sigma v_{\phi}\right)=0
\label{cont}
\end{eqnarray}
where $\Sigma$ is the surface density and $v_{R}$ and $v_{\phi}$ are the velocity components in the radial and azimuthal directions respectively. Since the disk is located in the $x-y$ plane, there are only two components for the Euler equation. These components can be , respectively, expressed as:
\begin{eqnarray}
\frac{\partial v_{R}}{\partial t}+v_{R} \frac{\partial v_{R}}{\partial R}+\frac{v_{\phi}}{R} \frac{\partial v_{R}}{\partial \phi}-\frac{v_{\phi}^{2}}{R}=- \frac{\partial}{\partial R}\left(\Phi+h\right)&\\
\frac{\partial v_{\phi}}{\partial t}+v_{R} \frac{\partial v_{\phi}}{\partial R}+\frac{v_{\phi}}{R} \frac{\partial v_{\phi}}{\partial \phi}+\frac{v_{\phi} v_{R}}{R}=- \frac{1}{R}\frac{\partial}{\partial\phi}\left(\Phi+h\right)
\label{euler}
\end{eqnarray}
where we have chosen a simple barotropic equation of state as $p=K \Sigma^{\gamma}$, where $K$ and $\gamma$ are constant real parameters. Also, $h$ is the specific enthalpy defined as:
\begin{eqnarray}
h= \int \frac{dp}{\Sigma} 
\end{eqnarray}
Now, let us perform the following perturbations: $\Sigma=\Sigma_{0}+\Sigma_{1}$, $v_{R}=v_{R0}+v_{R1}=v_{R1}$, $v_{\phi}=v_{\phi 0}+v_{\phi 1}$, $\Phi=\Phi_{0}+\Phi_{1}$ and $h=h_{0}+h_{1}$. We denote the equilibrium quantities with a ''$0$'' subscript and the perturbations with a ''$1$'' subscript. After linearization, the fluid equations (\ref{cont}) and (\ref{euler}) can be written as :  
\begin{eqnarray}
\frac{\partial \Sigma_{1}}{\partial t}+\frac{1}{R}\frac{\partial}{\partial R}\left(\Sigma_{0} R v_{R1}\right)+\Omega \frac{\partial \Sigma_{1}}{\partial \phi}+\frac{\Sigma_{0}}{R} \frac{\partial v_{\phi 1}}{\partial \phi}=0
\label{contin2}
\end{eqnarray}
\begin{eqnarray}
\frac{\partial v_{R1}}{\partial t}+\Omega \frac{\partial v_{R1}}{\partial \phi}-2\Omega v_{\phi 1}=-\frac{\partial}{\partial R}\left(\Phi_{1}+h_{1}\right)
\label{euler1}
\end{eqnarray}
\begin{eqnarray}
\frac{\partial v_{\phi 1}}{\partial t}+\Omega \frac{\partial v_{\phi 1}}{\partial \phi}+\frac{\kappa^{2}}{2\Omega} v_{R1}=-\frac{1}{R}\frac{\partial}{\partial \phi}\left(\Phi_{1}+h_{1}\right)
\label{euler2}
\end{eqnarray}
where $\Omega(R)$ is the circular frequency and the epicyclic frequency $\kappa$ is 
\begin{eqnarray}
\kappa^{2}(R)=R \frac{d\Omega^{2}}{dR}+4 \Omega^{2}
\end{eqnarray}
Also, the modified Poisson equation can be linearized as
\begin{eqnarray}
\begin{split}
 (\nabla^2-m_0^2)\nabla^2\Phi_1=&-4\pi G \alpha  \\& \times \left(m_0^2 \Sigma_1 \delta(z)-\frac{4}{3 }\nabla^2\Sigma_1 \delta(z)\right)
\label{po2}
\end{split}
\end{eqnarray}
Now, consider an arbitrary spiral surface density perturbation mode as follows
\begin{eqnarray}
\Sigma_1(R,\phi,t)=H(R)e^{i(F(R)+m \phi-\omega t)}
\label{tw}
\end{eqnarray}
where $\omega$ is the frequency of the mode, $H$ is a slowly varying function of $R$ and $F(R)$ is the \textit{shape function}. Also, $m$ is a positive integer, and the perturbation has $m$-fold rotational symmetry. The WKB approximation requires that $\frac{|k R|}{m}\gg 1$, where $k(R)=\frac{dF}{dR}$. In this approximation the pitch angle of the spiral arms is very small at any radius. Therefore, one can conclude that the spiral arms are tightly wound.

Let us find the gravitational potential of the tightly wound spiral perturbation (\ref{tw}) in the neighborhood of a point $(R_0,\phi_0)$. Using the properties of WKB approximation, one can expand $\Sigma_1$ as follows (for more detail see Binney \& Tremaine 2008)
  \begin{eqnarray}
  \Sigma_1(R,\phi,t)\sim \Sigma_a(R_0)e^{i(k(R_0)R-\omega t)}
  \label{swkb}
  \end{eqnarray}
where 
\begin{eqnarray}
\Sigma_a(R_0)=H(R_0)e^{i(F(R_0)+m \phi_0-k(R_0)R_0)}
\end{eqnarray}
the spiral perturbation (\ref{swkb}) is reminiscent of a plane wave with wavenumber $\mathbf{k} = k\mathbf{e}_R$, where $\mathbf{e}_R$ is the unit vector in the radial direction. The potential of a plane wave in a thin disk in the framework of the $f(R)$ gravity has been derived in Appendix \ref{app}. Using equation (\ref{finala}), the potential can be expressed as
\begin{eqnarray}
\Phi_1=-\frac{2\pi G \alpha}{|k|}\frac{m_0^2}{k^2+m_0^2}\Sigma_1
\label{po5}
\end{eqnarray}

Regarding the form of $\Sigma_1$ and $\Phi_1$, any solution of equations (\ref{contin2})-(\ref{euler2}) can be expressed as
\begin{eqnarray}
 v_{R1}(R,\phi,t)=v_{Ra}(R)e^{i(m \phi-\omega t)}&\\
v_{\phi1}(R,\phi,t)=v_{\phi a}(R)e^{i(m \phi-\omega t)}&\\
h_{1}(R,\phi,t)=h_a(R)e^{i(m \phi-\omega t)}
\end{eqnarray}
In Newtonian gravity, using the properties of the WKB approximation, equations (\ref{contin2})-(\ref{euler2}) take the following forms respectively (for more detail see Binney \& Tremaine 2008 )
\begin{eqnarray}
(m \Omega-\omega)\Sigma_a+k\Sigma_0 v_{Ra}=0
\label{conti3}
\end{eqnarray}
\begin{eqnarray}\label{euler3}
v_{Ra}=\frac{(m\Omega-\omega)k(\Phi_a+h_a)}{\Delta}&\\
v_{\phi a}=-\frac{2i B k(\Phi_a+h_a)}{\Delta}
\end{eqnarray}
where coefficient functions $\Delta$ and $B$ (the Oort constant of rotation) are:
\begin{eqnarray}
\Delta=\kappa^2-(m\Omega-\omega)^2 &\\B(R)=-\left(\Omega+\frac{1}{2}R\frac{d\Omega}{dR}\right)
\label{del}
\end{eqnarray}

It is important noting that mathematical form of the Euler equation in the weak field limit of $f(R)$ gravity is the same as the Newtonian gravity. The only difference between these theories appears in the way by which the disk responds to the perturbation. On the other hand, the Poisson equation determines the response or the behavior of the disk to the perturbation. Since, for obtaining equations (\ref{euler3}), the Poisson equation has not been used, so we can conclude that they are also true in the weak filed limit of $f(R)$ gravity. This is the case also for continuity equation (\ref{conti3}). Therefore, we skip the details of the calculations for deriving these equations, (\ref{conti3}) and (\ref{euler3}), and refer the reader to chapter 6 of Binney \& Tremaine 2008.

Now, we substitute $v_{Ra}$ from equation (\ref{euler3}) into equation (\ref{conti3}) and use equation (\ref{po5}). Also, taking into account that $h_a=v_s^2\frac{\Sigma_a}{\Sigma_0}$, we find the following dispersion relation
\begin{eqnarray}
(m\Omega-\omega)^2=\kappa^2-2\pi G |k|\Sigma_0\frac{\alpha m_0^2}{k^2+m_0^2}+k^2 v_s^2
\label{dis1}
\end{eqnarray}
We shall derive the local stability criterion from this dispersion relation. As expected, in the limit $m_0\rightarrow \infty$ and $\alpha\rightarrow 1$, the dispersion relation for WKB modes in Newtonian gravity can easily be recovered. In this paper, we restrict ourselves to the axisymmetric disturbances, i.e. $m=0$. Since the right-hand-side of (\ref{dis1}) is real, then the disk is stable against local axisymmetric disturbances if $\omega^2>0$ and unstable if $\omega^2<0$.

 It is worthwhile to mention that Toomre's criteria (\ref{Q1}) and (\ref{Q2}) have also been derived for axisymmetric perturbations. However, it cannot be concluded strongly that they are not applicable for local nonaxisymmetric stability. In fact, no general criterion for nonaxisymmetric stability is known in Newtonian gravity.

 Now, we consider the general case by adopting non-zero speed of sound. In this case, using the dispersion relation (\ref{dis1}), the stability criterion $\omega^2>0$ can be rewritten in the dimensionless form
 \begin{eqnarray}
 X^4+(1+\beta^2)X^2-\frac{2\beta\alpha}{Q_g}|X|+\beta^2>0
 \label{dis3}
 \end{eqnarray}
where 
\begin{eqnarray}
X=\frac{k}{m_0}~~~~,~~~~\beta=\frac{\kappa}{m_0 v_s}
\end{eqnarray}
%=========================================================================
\begin{figure}
\begin{center}
\resizebox{8cm}{!}{\includegraphics{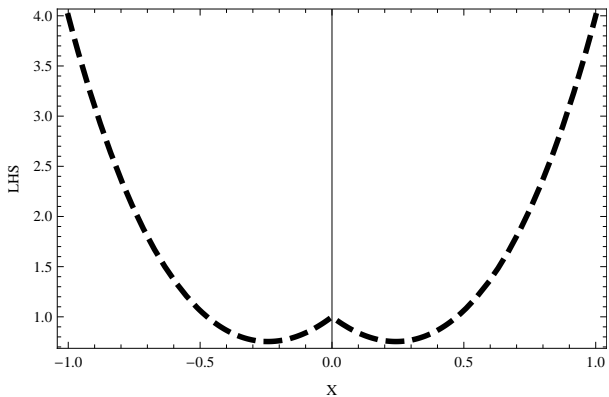}} \hfill
%\parbox[b]{55mm}
\caption{Schematic behavior of the LHS of equation (\ref{dis3}) with respect to $X$.}
\label{1}
\end{center}
\end{figure}
%======================================================================

 The schematic behavior of the left-hand side (LHS) of equation (\ref{dis3}) is shown in Figure \ref{1}. It is clear from this figure that, if the minimum value of the LHS is positive, then the condition $\omega^2>0$ will hold for all wavenumbers and consequently throughout the region that we explore, all solutions that we will find represent stable modes. In other words, the disk is stable against all axisymmetric tightly wound perturbations, if the minimum value of the LHS is positive. It is easy to check that the minimum value of the LHS occurs at $X_{min}$ given by
\begin{eqnarray}
X_{min}=\pm\left[-\frac{1+\beta^2}{3}A^{1/3}+\frac{1}{2}A^{-1/3}\right]
\end{eqnarray}
where $A$ is defined as 
\begin{eqnarray}
A=\frac{Q_g}{2\beta\alpha\left(1+\sqrt{1+\frac{2(1+\beta^2)^3Q_g^2}{27\beta^2\alpha^2}}\right)}
\end{eqnarray}
The location of this point varies as the parameters $\alpha$ and $\beta$ change. Finally the local stability criterion can be expressed as
 \begin{eqnarray}
 Q_g>\frac{2\beta\alpha|X_{min}|}{X_{min}^4+(1+\beta^2)X_{min}^2+\beta^2}
 \label{dis4}
 \end{eqnarray}
This is the main result of this section. In fact, equation (\ref{dis4}) is the generalized version of the Toomre's local stability criterion for a fluid disk, and should be compared with the standard case given by (\ref{Q1}). Since $X_{min}$ contains $Q_g$, criterion (\ref{dis4}) is very complicated than the standard Toomres criterion (i.e. $Q_g>1$). However, it is straightforward to show that in the limit $\beta\rightarrow 0$ and $\alpha\rightarrow 1$, the right-hand side (RHS) of equation (\ref{dis4}) equals $1$, and the standard Toomre's criterion is recovered. 

\begin{figure}
\begin{center}
\resizebox{8cm}{!}{\includegraphics{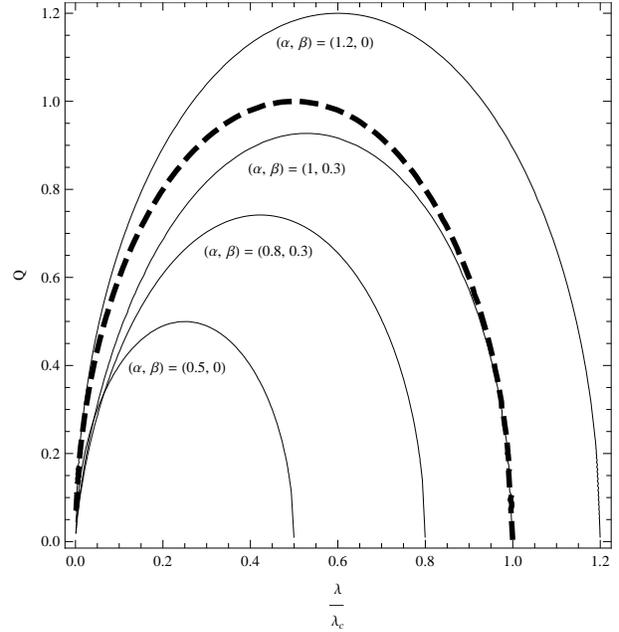}} \hfill
%\parbox[b]{55mm}
\caption{The boundary of stable and unstable axisymmetric WKB disturbances in a fluid disk for different values of $\alpha$ and $\beta$. The dashed curve corresponds to Newtonian gravity where $\alpha=1$ and $\beta=0$.}\label{bound2}
\end{center}
\end{figure}

It is also instructive to plot the neutral stability curves $\omega=0$ for axisymmetric tightly wound waves in a fluid disk. These curves specify the boundary of the stable and unstable modes/waves. To do so, let us introduce a critical wavelength $\lambda_{crit}=2\pi/k_{crit}$. Using this definition and the dispersion relation (\ref{dis1}), the line separating stable and unstable modes is given by
\begin{eqnarray}
Q_g(y)=\alpha\sqrt{\frac{4\alpha y}{1+\left(\frac{\beta Q_g}{2 y\alpha}\right)^2}-4y^2}
\label{bound}
\end{eqnarray} 
where $y=\lambda/\lambda_{crit}$ and $\lambda=2\pi/k$. We have plotted the neutral curve for various values of $\alpha$ and $\beta$ in Figure \ref{bound2}. The dashed curves corresponds to Newtonian gravity where $(\alpha,\beta)=(1,0)$. 

Let us first consider the other curves with $\beta=0$, i.e. curves corresponding to $(\alpha,\beta)=(1.2,0)$ and $(\alpha,\beta)=(0.5,0)$. In this case one can readily show that the Toomre's criterion (\ref{dis4}) is $Q_g>\alpha$. Consequently, if $\alpha>1$ (as in the case ($\alpha,\beta)=(1.2,0)$) then a lager $Q_g$ parameter compared with Newtonian gravity is needed to overcome the local gravitational instability. Keeping in mind the physical interpretation of the Jeans instability, this means that the disk needs more gas pressure than it would in Newtonian gravity in order to overcome the gravitational collapse. On the other hand, if $\alpha<1$ then smaller gas pressure is needed for preventing the collapse. 

For the general case $\beta\neq 0$ see the curves $(\alpha,\beta)=(1,0.3)$ and $(\alpha,\beta)=(0.8,0.3)$ in Figure \ref{bound2}. In this case the neutral curves corresponding to $\beta\neq 0$ lie below the dashed curve. This can not be explained with the above mentioned simple interpretation and one should consider it with more care. In order to describe this result, for the sake of simplicity and with no loss of generality we assume that $\alpha=1$ and $\beta\ll 1$. We shall see that in reality $\beta$ is small and our assumption here is a suitable one. Therefor, the RHS of the criterion (\ref{dis4}) can be expanded in terms of $\beta$. Consequently, we have
\begin{eqnarray}
Q_g>1-\beta^2-\frac{3}{2}\beta^4+O(\beta^5)
\label{req}
\end{eqnarray}
Thus in the case $\alpha=1$, although the gravitational force is supposed to be stronger than the Newtonian case, the required value of $Q_g$ for local stability is always smaller than that of the Newtonian gravity. This situation is reminiscent of the difference between Jeans mass in modified gravity and  Newtonian gravity. In fact in modified theories of gravity for which the gravitational force is stronger than that of Newtonian gravity, the Jeans mass is smaller than the standard case, for example see \citep{us2,maria}.

Therefore, comparing the two cases $\beta=0,~ \alpha>1$ and $\alpha=1, ~ \beta>0$ (note that in both cases the gravitational force is stronger than the Newtonian gravitational force), we can conclude that in theories where the gravitational force in the weak field limit is stronger than that of Newtonian gravity, the required value of Toomre's parameter $Q_g$ for preventing gravitational collapse can be larger or smaller than the corresponding value in Newtonian gravity. In other words, $Q_g$ in these theories is not necessarily larger than that of Newtonian gravity.

\begin{figure}
\begin{center}
\resizebox{8cm}{!}{\includegraphics{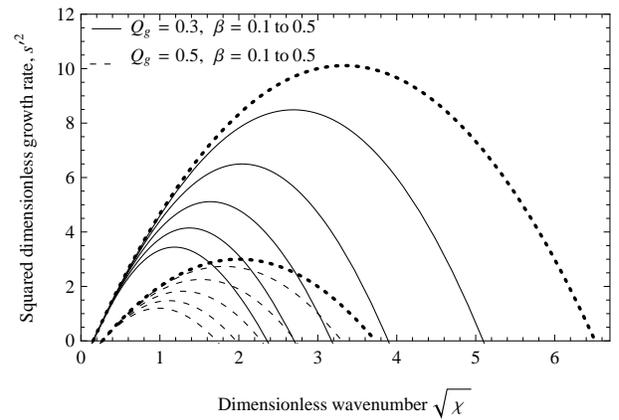}} \hfill
%\parbox[b]{55mm}
\caption{Solutions to the dispersion relation (\ref{dis1}) between squared growth rate ($s'^2$) and dimensionless wavenumber ($k v_s/\kappa$). It has been assumed that $\alpha=1$. Large $\beta$ leads to greater instability at small wavenumber.}\label{new2}
\end{center}
\end{figure}

In Figure \ref{new2} we show the fluid disk dispersion relation for growing perturbations for various values of $Q_g$ and $\beta$. The solid curves correspond to $Q_g=0.3$ and $\beta=0.1$ to $0.5$. Furthermore, the dashed curves correspond to $Q_g=0.5$ and $\beta=0.1$ to $0.5$. Also, the dotted curves ($\beta=0$) are the corresponding Newtonian dispersion relations for $Q_g=0.3$ and $Q_g=0.5$. As expected, the growth rates tends to increase with decreasing $Q_g$. Also, it is clear form Figure \ref{new2} that the effect of parameter $\beta$ in equation (\ref{dis1}) is to stretch the range of instability to small wavenumber while larger $\beta$ leads to smaller growth rate.

%=========================================================================
%=========================================================================
%=========================================================================

\subsection{Self-gravitating stellar disk}\label{self stellar}
Similar to the case of the fluid disk, it is straightforward to find the dispersion relation of tightly wound waves in a stellar disk. Since most of the calculations are similar to that of the fluid case, we have briefly derived the dispersion relation in Appendix \ref{app2}. The final result is given by equation (\ref{dis5}). In the case of axisymmetric perturbations, we may write
\begin{eqnarray}
\omega^2=\kappa^2-2\pi G |k|\Sigma_0\frac{\alpha m_0^2}{k^2+m_0^2}\mathfrak{F}(\frac{\omega}{\kappa},\frac{k^2 \sigma_R^2}{\kappa^2})
\label{dis6}
\end{eqnarray}
%====================================================================
\begin{figure}
\begin{center}
\resizebox{8cm}{!}{\includegraphics{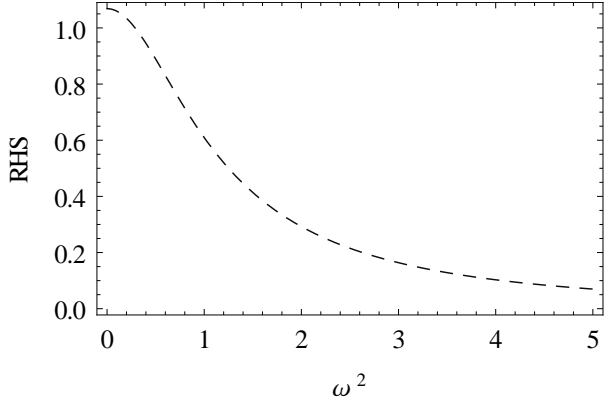}} \hfill
%\parbox[b]{55mm}
\caption{equation (\ref{dis8}) in terms of $\omega^2$.}\label{chap}
\end{center}
\end{figure}
%====================================================================
\begin{figure}
\begin{center}
\resizebox{8cm}{!}{\includegraphics{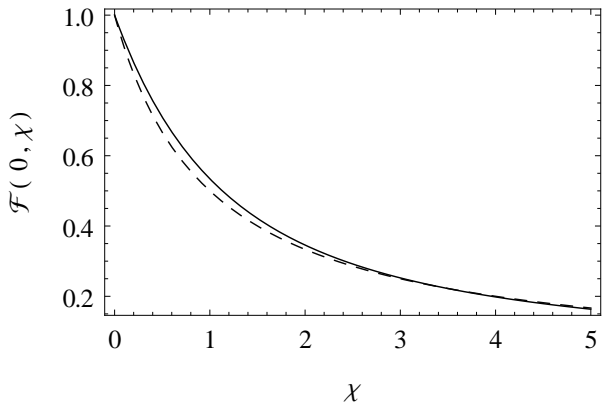}} \hfill
%\parbox[b]{55mm}
\caption{The solid curve is the exact curve for $\mathfrak{F}(0,\chi)$, and the dashed curve is the approximate one, i.e. equation (\ref{recap}).}\label{reduc}
\end{center}
\end{figure}
%=========================================================================

For a given mode of perturbation, if $\omega^2<0$, then the mode is unstable. Therefore, one may expect that the system is stable against all axisymmetric disturbances if equation (\ref{dis6}) does not have a solution with $\omega^2<0$. In order to consider this expectation, we write down equation (\ref{dis6}) as follows
\begin{eqnarray}
1=\frac{2\pi G |k|\Sigma_0 \alpha m_0^2}{(k^2+m_0^2)(\kappa^2-\omega^2)}\mathfrak{F}(\frac{\omega}{\kappa},\frac{k^2 \sigma_R^2}{\kappa^2})
\label{dis7}
\end{eqnarray}
Now, if $\omega^2<0$, i.e. $\omega=i \gamma$ where $\gamma$ is a real number, then the right-hand side (RHS) of equation (\ref{dis7}) takes the form
\begin{eqnarray}
\frac{2\pi G |k| \alpha m_0^2 \Sigma_0\kappa^{-2}
}{(k^2+m_0^2)\sinh \pi s'}\int_0^{\pi}e^{-\chi(1+\cos\tau)}\sinh s'\tau \sin \tau d\tau
\label{dis8}
\end{eqnarray}
where $s'=i s=\gamma/\kappa$. Also we have used equation (\ref{shap}) for the shape function. The schematic behavior of (\ref{dis8}) in terms of $\omega^2$ is illustrated in Figure \ref{chap}. From this figure we see that the maximum value occurs at $\omega^2=0$. Therefore, if the RHS of equation (\ref{dis6}) for $\omega^2=0$ is smaller than $1$, then there is no tightly wound spiral mode with negative $\omega^2$ which satisfies the dispersion relation (\ref{dis7}). Consequently, the self-gravitating disk will be stable to all WKB perturbations. Therefore, the criterion for stability to all wavelengths can be written down as
 \begin{eqnarray}
 \frac{2\pi G |k|\Sigma_0 \alpha m_0^2}{(k^2+m_0^2)\kappa^2}\mathfrak{F}(0,\frac{k^2 \sigma_R^2}{\kappa^2})<1
 \label{dis9}
 \end{eqnarray}
 this equation can be represented in a more simplified form as
 \begin{eqnarray}
\frac{\sigma_R \kappa}{2\pi G \Sigma_0\alpha}>\frac{1}{(1+\beta^2\chi)}\frac{\left(1-e^{-\chi}I_0(\chi)\right)}{\sqrt{\chi}}
 \label{dis10}
 \end{eqnarray}
 where $I_0$ is the modified Bessel function of order $0$. Also we have used the following identity
 \begin{eqnarray}
 \mathfrak{F}(0,\chi)=\frac{1}{\chi}\left(1-e^{-\chi}I_0(\chi)\right)
 \end{eqnarray}
 Equation (\ref{dis10}) is the main result of this section. In fact, at every location $(R_0,\phi_0)$ on the disk, this inequality should be satisfied in order to have local stability to all spiral axisymmetric tightly wound perturbations. As $\alpha\rightarrow 1$ and $\beta\rightarrow 0$, this criterion approaches to the so-called Toomre's local stability criterion for stellar disk given by (\ref{Q2}), again as expected. In Newtonian gravity the RHS of (\ref{dis10}) is always larger than that of metric $f(R)$ gravity. This means that in the context of $f(R)$ gravity smaller value of Toomre's parameter is required for local stability of rotating disks. As we discussed in the previous subsection, this result also is the case for a fluid disk.

Although equation (\ref{dis10}) is our final stability criterion for every perturbation with wavenumber $k$, if we find the maximum value of the RHS of equation (\ref{dis10}), then we will find a criterion which guarantees the stability at every location on the disk. Even in Newtonian gravity (where $\beta=0$), one has to use numerical analysis to find the maximum value of the RHS (\ref{dis10}). Therefore, we need the numeric value of $\beta$ to find required criterion. However, using an excellent approximation for reduction factor \citep{elmegreen}, we can analytically find the maximum value of RHS of (\ref{dis10}). The above mentioned approximation is
\begin{eqnarray}
\mathfrak{F}(0,\chi)\simeq \frac{1}{1+\chi}
\label{recap}
\end{eqnarray}
In the Figure \ref{reduc}, the solid line is the exact value of $\mathfrak{F}$ and the dashed line corresponds the approximate value. As it is clear from the figure, the deviation between the two functions is small. Therefore, with the aid of this approximation, we rewrite (\ref{dis10}) as follows
 \begin{eqnarray}
\frac{\sigma_R \kappa}{2\pi G \Sigma_0\alpha}>\frac{\sqrt{\chi}}{(1+\beta^2\chi)(1+\chi)}
 \label{dis100}
 \end{eqnarray}
the maximum value of the RHS of this inequality occurs
 \begin{eqnarray}
 \chi=\frac{-1-\beta^2+\sqrt{1+14\beta^2+\beta^4}}{6\beta^2}
 \end{eqnarray}
stability criterion takes the following form
 \begin{eqnarray}
Q_s>\frac{6\sqrt{6}\alpha\sqrt{g(\beta)-(1+\beta^2)}}{(g(\beta)+5-\beta^2)(g(\beta)+5\beta^2-1)}
 \label{dis101}
 \end{eqnarray} 
 where $g(\beta)=\sqrt{1+14\beta^2+\beta^4}$. The criterion (\ref{dis101}) is the final result of this section. As expected, the free parameters of the theory, i.e. $\alpha$ and $m_0$, appear in the stability criterion. If we use the same approximation in Newtonian gravity, then we will get the following criterion
  \begin{eqnarray}
\frac{\sigma_R \kappa}{3.14 G \Sigma_0}>1
 \label{dis102}
 \end{eqnarray} 
this criterion has a little difference with the exact criterion (\ref{Q2}). However let us write (\ref{dis102}) as $Q_s>1$. As we discussed previously, we expect that $\beta$ be small in real situations, so the stability criterion can be expanded with respect to $\beta$ as
 \begin{eqnarray}
 Q_s>\alpha-\alpha \beta^2+3\alpha \beta^4+O(\beta^5)
\end{eqnarray}  
this criterion can be compared with the corresponding criterion for a fluid disk (\ref{req}). The similarity between the stability criterion of fluid and stellar disk is obvious. Our discussion in the last paragraph of section \ref{self fluid} considering the parameter $Q_g$ for a fluid disk, also holds here.

\begin{figure}
\begin{center}
\resizebox{8cm}{!}{\includegraphics{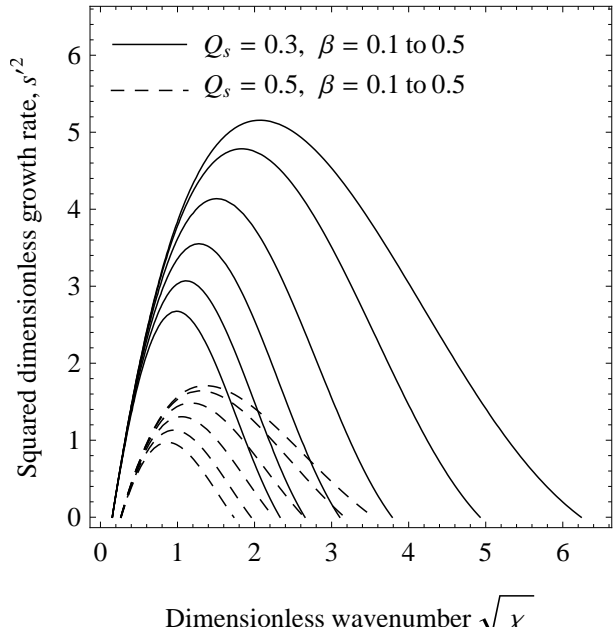}} \hfill
%\parbox[b]{55mm}
\caption{Solutions to the dispersion relation (\ref{dis6}) between squared growth rate ($s'^2$) and dimensionless wavenumber ($k v_s/\kappa$). It has been assumed that $\alpha=1$. Large $\beta$ leads to greater instability at small wavenumber.}\label{new1}
\end{center}
\end{figure}

In order to make a comparison between the growth rates in fluid and stellar disks, we have shown the stellar disk dispersion relation (\ref{dis6}) in Figure \ref{new1}. The solid curves correspond to $Q_s=0.3$ and $\beta=0.1$ to $0.5$. The dashed curves correspond to $Q_s=0.5$ and $\beta=0.1$ to $0.5$. Also, the dotted curves are the corresponding Newtonian dispersion relations for $Q_s=0.3$ and $Q_s=0.5$ where $\beta=0$. As for a fluid disk, the growth rates increase with decreasing parameter $Q_s$, and the effect of parameter $\beta$ in equation (\ref{dis6}) is to stretch the range of instability to small wavenumber. Also larger $\beta$ leads to smaller growth rate. Comparing Figures \ref{new2} and \ref{new1}, one can conclude that the the growth rate of perturbation in the stellar disk is smaller than that of the fluid disk. In other words, if we consider two patches with the same Toomre's parameter, one in a fluid disk and one in a stellar disk, the the growth rate in the patch on the stellar disk will be smaller than that of the fluid disk. 
%=========================================================================
\section{discussion and conclusion}\label{DC}
In this paper, the local stability of self-gravitating fluid and stellar disks has been investigated in the framework of metric $f(R)$ gravity, for which the function  $f(R)$ can be expanded as a power series in terms of Ricci scalar $R$, see equation (\ref{model}). Thus the results of this paper are applicable only for these models of $f(R)$. The dispersion relation for the propagation of tightly wound spiral waves has been derived analytically for both fluid and stellar disks. The results are given by  equation (\ref{dis1}) for fluid disk and (\ref{dis6}) for the stellar disk.

Also, using the above mentioned dispersion relations, the Toomre's local stability criterion has been derived in the framework of metric $f(R)$ gravity. The results are given by equations (\ref{dis4}) and (\ref{dis10}). An important ingredient of the present study is its possible application of these criteria in N-body simulations of the disk galaxies in the context of $f(R)$ gravity. It is worth mentioning that the Toomre's criterion for the fluid disk (\ref{dis4}) is complicated than the standard case (\ref{Q1}). On the other hand, the generalized Toomre's criterion for the stellar disk is not so different from its corresponding criterion in Newtonian gravity (\ref{Q2}).

As we have discussed, the magnitude of $m_0$ determines the magnitude of deviation between local stability criteria of $f(R)$ gravity and GR. Therefore, it is necessary to mention that the mass of the effective scalar degree of freedom, $m_0$, may depend upon the density of its environment via the so-called \textit{chameleon mechanism} \citep{khourya,khouryb}. Obviously, a more careful study is still in order to take into account the chameleon mechanism. Studying this subject is left as a subject of future study.

\acknowledgments 
This work is supported by Ferdowsi University of
Mashhad under Grant No. 100836 (25/05/1393).
\makeatletter
\let\clear@thebibliography@page=\relax
\makeatother

\appendix
\section{Potential of a WKB wave in $f(R)$ gravity}\label{app}
As we mentioned before, locally, a tightly wound spiral wave can be considered as a plan wave. The reason for this similarity is that the pitch angle is very small at every location for tightly wound density waves. It is important to remember that the surface density corresponding to a tightly wound spiral wave is given by (\ref{swkb}). In order to find the gravitational potential of this perturbation, without loss of generality, we choose the $x$ axis to be parallel to $\textbf{k}(R_0)$. Then we use the modified Poisson equation (\ref{po}) to find the gravitational potential. To do so, we guess the solution to (\ref{po}) as follows 
\begin{eqnarray}
\Phi_{1}(x,y,z,t)=\lambda~ e^{i (k x-\omega t)-|\epsilon z|}
\label{po3}
\end{eqnarray}
where
$\lambda$
and
$\epsilon$
are arbitrary real constants. The disk is assumed to be thin and so there is no matter outside ($z\neq 0$) the disk. Therefore outside the disk the modified Poisson equation (\ref{po}) can be written down as 
$\nabla^{4}\Phi_1-m_0^2\nabla^2\Phi_1=0$. Substituting the potential (\ref{po3}) into this equation, one can easily verify that 
$\epsilon=\pm k$
. 
On the other hand, since matter is located at $z=0$, derivative of
$\Phi_1$ with respect to $z$ is not continuous. In order to fix the parameter $\lambda$, we integrate equation (\ref{po2}) with respect to $z$ in the interval $z=-\xi$ to $z=+\xi$, where $\xi$ is a positive constant,
and then let $\xi\rightarrow 0$. Therefore, the LHS of (\ref{po2}) gives
\begin{eqnarray}
\lim _{\xi\rightarrow 0 }\int_{-\xi}^{+\xi}dz(\nabla^{4}\Phi_1-m_0^2\nabla^2\Phi_1)=\lim _{\xi\rightarrow 0 }\int_{-\xi}^{+\xi}dz \left( \frac{\partial^4 \Phi_1}{\partial z^4}+2\frac{\partial^4 \Phi_1}{\partial x^2 \partial z^2}-m_0^2 \frac{\partial^2 \Phi_1}{\partial z^2}\right)=&\\
\lim _{\xi\rightarrow 0 }\left( \frac{\partial^3 \Phi_1}{\partial z^3}+2\frac{\partial^3 \Phi_1}{\partial x^2 \partial z}- m_0^2 \frac{\partial \Phi_1}{\partial z}\right)=&2|k|\lambda(k^2+m_0^2)e^{i(k x -\omega t)}\nonumber
\label{atwp}
\end{eqnarray}
note that we have not written those partial derivatives of $\Phi_1$ which are continuous at $z=0$, because their integral with respect to $z$ in the above mentioned interval is zero. The same procedure for the RHS of (\ref{po2}) gives
\begin{eqnarray}
-4\pi G \alpha m_0^2\lim _{\xi\rightarrow 0 }\int_{-\xi}^{+\xi}dz\left( \Sigma_1 \delta(z)-\frac{4}{3m_0^2 }\nabla^2[\Sigma_1 \delta(z)]\right)
\label{a}
\end{eqnarray}
Taking into account that the tightly wound density wave is $\Sigma_1=\Sigma_a e^{i(k x-\omega t)}$, and using the properties of the Dirac delta function, one can show that the second integral in equation (\ref{a}) vanishes. Therefore, equation (\ref{a}) can be simplified as follows
\begin{eqnarray}
-4\pi G \alpha m_0^2 \Sigma_a e^{i(k x-\omega t)}
\label{a2}
\end{eqnarray}
finally, by equating equation (\ref{atwp}) to (\ref{a2}), we find the relation between $\lambda$ and $\Sigma_a$ as
\begin{eqnarray}
\lambda=-\frac{2\pi G\alpha}{|k|}\frac{m_0^2}{k^2+m_0^2}\Sigma_a
\label{finala}
\end{eqnarray}
as expected, in the limit $m_0^2\rightarrow \infty$ and $\alpha\rightarrow 1$, we recover the potential of the tightly wound spiral in  Newtonian gravity. 

\section{Dispersion relation for stellar disk in $f(R)$ gravity}\label{app2}
In order to study the dynamics of a self-gravitating stellar disk in the framework of metric $f(R)$ gravity, we need the modified Poisson equation (\ref{po}) and the collisionless Boltzmann equation (\ref{bolt}). Using (\ref{bolt}), one can readily derive the Euler and continuity equations, respectively, as follows
\begin{eqnarray}\label{B1}
\frac{\partial \bar{v}_j}{\partial t}+\bar{v}_i\frac{\partial \bar{v}_j}{\partial x^i}=-\left[\frac{\partial \Phi}{\partial x^i}+\frac{1}{\Sigma}\frac{\partial}{\partial x^i}(\sigma_{ij}^2\Sigma)\right]
\end{eqnarray}
\begin{eqnarray}\label{B2}
\frac{\partial \Sigma}{\partial t}+\frac{\partial}{\partial x^i}(\Sigma \bar{v}_i)=0
\end{eqnarray}
where
\begin{eqnarray}
\sigma_{ij}^2=\overline{v_i v}_j-\bar{v}_i\bar{v}_j~~~,~~~\bar{v}_i=\frac{1}{\Sigma}\int f v_i d^2x
\end{eqnarray}
and $f$ is the distribution function. Combining equations (\ref{B1}) and (\ref{B2}) with (\ref{po}) and using the same general technique that we used in section \ref{self fluid} to derive the dispersion relation for a fluid disk, we find
\begin{eqnarray}
\bar{v}_{Ra}=\frac{m\Omega-\omega}{\Delta}k\Phi_a \mathfrak{F}
\end{eqnarray}
 where $\mathfrak{F}$ is the \textit{reduction factor}. It is the factor by which the response of a stellar disk to an imposed potential is reduced below that of a cold disk (see \citep{binney} for more detail about this factor and the reason for which it appears in the linearized equations), and $\Delta$ has been defined in (\ref{del}). Also, one can show that equations (\ref{conti3}) and (\ref{po5}) are also applicable for describing a tightly wound density wave in the stellar disk. Therefore, one can find after some algebra that the dispersion relation is
\begin{eqnarray}
(m \Omega-\omega)^2=\kappa^2-2\pi G |k|\frac{\alpha m_0^2}{k^2+m_0^2}\mathfrak{F}(s,\chi)
\label{dis5}
\end{eqnarray}
where $s$ and $\chi$ are defined as
\begin{eqnarray}
s=\frac{\omega-m\Omega}{\kappa}~~~,~~~\chi=\left(\frac{k\sigma_R}{\kappa}\right)^2
\end{eqnarray}
As in Newtonian gravity, the difference between the dispersion relation of a fluid disk and a stellar one manifest itself in the appearance of the reduction factor in the dispersion relation of the stellar disk. The reduction factor in Newtonian gravity is given by (see \citep{binney} and references therein)
\begin{eqnarray}
\mathfrak{F}(s,\chi)=\frac{1-s^2}{\sin\pi s}\int_{0}^{\pi}e^{-\chi(1+\cos\tau)}\sin s\tau ~\sin\tau d\tau
\label{shap}
\end{eqnarray} However, a natural question arises here: Does the reduction factor $\mathfrak{F}$ in metric $f(R)$ gravity and Newtonian gravity coincide? It is worth mentioning that in obtaining the reduction factor, the Poisson equation is not used (see Appendix $K$ of \citep{binney} for a full derivation of the reduction factor). On the other hand, at least at the level of field equations, the only difference between Newtonian gravity and the weak field limit of metric $f(R)$ gravity appears in their Poisson equation. Therefore, we expect that the reduction factor must be the same in both $f(R)$ and Newtonian gravity. However, we need to treat this conclusion more carefully. In fact for obtaining the reduction factor some assumptions have been done, and one needs to check their consistency in metric $f(R)$ gravity. This factor has been derived implementing three main assumptions. 1) the given perturbation lies in the WKB approximation. 2) The stellar orbit of stars can be described by the epicycle approximation in which the orbits are nearly circular. 3) The disk is thin and its distribution function is the so-called \textit{Schwarzschild} distribution function \citep{binney}. The first assumption is  satisfied in this work because we are studying tightly wound waves. The main consequence of the second assumption which is necessary for deriving the reduction factor is the relation between $\sigma_R$ and $\sigma_{\phi}$ in the epicycle approximation as \citep{binney} 
\begin{eqnarray}
\frac{\sigma_{\phi}^2}{\sigma_R^2}\simeq\frac{1}{2}\left(1+\frac{d R\Omega(R)}{d \ln R }\right)
\end{eqnarray}
this equation is derived only using the kinematic properties of the nearly circular motion and does not depend on the gravitational theory. Thus this relation can also be used in modified gravity theories including $f(R)$ gravity. Finally, for the third assumption it is just needed to note that, as we showed in subsection \ref{bolbol}, the mathematical form of the collisionless Boltzmann equation in $f(R)$ and Newtonian gravity is the same. Consequently, one can readily verify that the Schwarzschild distribution function is also a solution for (\ref{bolt}). Therefore, we can be sure the reduction factor (\ref{shap}) is also true in metric $f(R)$ gravity.
\end{document}